\documentclass[aps,prd,
groupedaddress,amssymb,
showpacs,nofootinbib]{revtex4}

\usepackage{amsmath,amssymb}

\usepackage[usenames]{color}

\usepackage{graphicx}
\usepackage{mathptmx}
\usepackage{bm}
\usepackage{times}

\newcommand{\beq}{\begin{eqnarray}}
\newcommand{\eeq}{\end{eqnarray}}

\begin{document}

\onecolumngrid



\title{Summing Planar Diagrams by an Integrable Bootstrap}



\author{Peter \surname{Orland}}

\email{orland@nbi.dk}


\affiliation{1. Baruch College, The 
City University of New York, 17 Lexington Avenue, 
New 
York, NY 10010, U.S.A. }

\affiliation{2. The Graduate School and University Center, The City University of New York, 365 Fifth Avenue,
New York, NY 10016, U.S.A.}

\affiliation{3. The Niels Bohr Institute, The Niels Bohr International Academy, Blegdamsvej 17, DK-2100, Copenhagen {\O}, Denmark}


\begin{abstract}

Correlation functions of matrix-valued fields are not generally known for massive 
renormalized field theories. We find
the large-$N$ limit of form factors of the
$(1+1)$-dimensional sigma model with ${\rm SU}(N)\times {\rm SU}(N)$ symmetry. These form factors give a correction to the free-field approximation for the $N=\infty$ Wightman function. The method 
is a combination of the $1/N$-expansion of the {\em S} matrix and Smirnov's form-factor axioms. We expand the
renormalized field in terms of a free massive Bosonic 
field as $N\rightarrow \infty$.

\end{abstract}


\maketitle

\section{Introduction}
\setcounter{equation}{0}
\renewcommand{\theequation}{1.\arabic{equation}}

The planarity of Feynman diagrams in the large-$N$
limit of matrix theories \cite{'t} has 
convinced many people that this limit is solvable. Unfortunately, little is 
known with precision about the $1/N$-expansion of 
($N\times N$)-matrix-valued field theories 
with {\em propagating degrees of 
freedom} ({\em i.e.} particles). Aside from 
maximally-supersymmetric, conformal-invariant theories, the only exceptions are 
($1+1$)-dimensional quantum chromodynamics \cite{'t-1+1} and string models with Chan-Paton factors \cite{GT}. Massive matrix-field theories are not solvable by straightforward 
saddle-point approaches. The saddle-point
method works only for field theories
whose $N=\infty$ diagrams are not just planar, but linear. In this paper, we make 
some progress by
melding the large-$N$ expansion with the form-factor bootstrap. Perhaps our results
will point to the solution of 
the planar limit in situations where this bootstrap does not work.

The {\em S} matrix
of the (1+1)-dimensional nonlinear sigma model with ${\rm SU}(N) \times {\rm SU}(N)$ symmetry is known. Unfortunately, its form factors are not, with the notable exception of the model with
${\rm SU}(2)\times {\rm SU}(2)\simeq {\rm O}(4)$ symmetry \cite{KW}. We 
study here the leading $1/N$-expansion of the form factors of the 
${\rm SU}(N) \times {\rm SU}(N)$-symmetric sigma model, also
known as the principal chiral model. The bare 
field is a matrix $U(x)$, lying in the fundamental representation of SU($N$), where $x^{0}$ and $x^{1}$ are the time and space coordinates, respectively, of (1+1)-dimensional Minkowski 
space-time. The action is
\beq
S=\frac{N}{2g_{0}^{2}}\int d^{2}x \;\eta^{\mu\nu}\;{\rm Tr}\,\partial_{\mu}U(x)^{\dagger}\partial_{\nu}
U(x),
\label{action}
\eeq
where $\mu, \nu=0,1$, $U(x)\in {\rm SU}(N)$ (that is, $U(x)$ is an $N\times N$ unitary matrix of determinant one), and the metric is that of flat Minkowski space, $\eta^{00}=1$, $\eta^{11}=-1$, $\eta^{01}=\eta^{10}=0$. The action does not change under the global transformation 
$U(x)\rightarrow V_{L}U(x)V_{R}$,  for two constant matrices $V_{L}, \,V_{R}\in {\rm SU}(N)$. We do not consider the addition of a Wess-Zumino-Witten term to this action. The sigma model is asymptotically free. All the evidence 
indicates that the Hamiltonian spectrum 
has a mass gap $m_{1}$, though no rigorous proof exists.

We study here the one-particle and three-particle form factors of the renormalized field operator 
$\Phi(x)$ (there are no two-particle form factors for $N>3$). This field may be expressed in a theory with ultraviolet cut-off $\Lambda$ as
\beq
\Phi(x)={\mathcal Z}(g_{0},\Lambda)^{-1/2}U(x), \label{phi}
\eeq 
where $g_{0}$ is the coupling. The renormalization factor 
${\mathcal Z}(g_{0}(\Lambda),\Lambda)$ vanishes in the limit $\Lambda\rightarrow \infty$, where the running coupling
$g_{0}(\Lambda)$ is defined so that the mass gap
$m_{1}(g_{0}(\Lambda), \Lambda)$ is independent of $\Lambda$. 



The {\em S} matrix of the principal chiral model has been found using the integrable bootstrap \cite{AALS}, \cite{Wiegmann} and a subtle Bethe {\em Ansatz} 
argument \cite{Pol-Wieg}. The essential
ideas of the former approach begin from a general classification of U($N$)-symmetric 
S-matrices
for vector particles \cite{BKKW}. One such {\em S} matrix has no backward 
scattering \cite{KS}, hence the effective
symmetry is SU($N$). The tensor product of two of these vector-particle 
S-matrices  
yields the general {\em S} matrix with  ${\rm SU}(N)\times {\rm SU}(N)$ symmetry, up to a CDD 
factor. The 
requirement of a sine formula for bound-state masses (which follows from relativistic kinematics
\cite{STW}) restricts the form of the CDD factor. 

In this paper, we combine the $1/N$-expansion of the {\em S} matrix \cite{Wiegmann} with Smirnov's axioms \cite{Smirnov}, to obtain the three-particle form 
factors of the renormalized field operator $\Phi(x)$. The 
LSZ reduction formula is used to
fix the overall normalization \cite{Ising}. 

There is an obvious advantage using the $1/N$-expansion to study correlation functions. Field theories with unitary symmetry have both fundamental or {\em elementary} particles and bound states. Particle masses are given by the sine formula mentioned above:
\beq
m_{r}=m_{1}\frac{\sin \frac{\pi r}{N}}{\sin\frac{\pi}{N}},\; \; r=1,\dots,N-1, \label{masses}
\eeq
where each choice of $r>1$ corresponds to a bound state of $r$ elementary particles. These 
bound states reveal themselves as poles in {\em S} matrix elements. Particles with $r>1$ make 
the determination of form factors difficult, though progress has been made \cite{BFK}. The picture simplifies dramatically
as $N\rightarrow \infty$, because the 
binding energy per particle number vanishes. The asymptotic states of the {\em S} matrix, with $r$ or 
$N-r$ finite, consist only of  $r=1$ particles and $r=N-1$ antiparticles, to any finite order of $1/N$. There are, however, bound states of infinite numbers of elementary particles, which correspond to keeping $r/N=\rho$ fixed, as $N\rightarrow \infty$ \cite{Herbert}. These bound states of infinitely many particles have mass $\approx 
N m_{1}(\sin \rho)/\pi$, which becomes infinite in 
the 't~Hooft limit, with $m_{1}$ fixed. There are continuously many such bound states, so their measure of integration must also be considered. We believe, however, that such bound states do not 
contribute to the $N\rightarrow \infty$ Wightman correlation function; they would produce 
unphysical cuts in momentum space. In an alternative large-$N$ limit (not 
the 't~Hooft limit, which we examine here), with $m_{1}/\sin\frac{\pi}{N}\approx Nm_{1}/\pi$ fixed, the parameter $r/N$ becomes continuous, playing the role of a 
third space-time dimension \cite{FKW}.

The main drawback of our approach is that bound-state corrections are not 
analytic in powers of $1/N$. In our view, this is outweighed by the simplicity of the
form-factor bootstrap in the planar limit.

Our interest in this problem began with applications of exact S-matrices and form factors of the 
SU($N$) sigma model to 
$(2+1)$-dimensional SU($N$) gauge theories \cite{Latt2+1}. The quark-antiquark potential 
\cite{FFgauge} and the gluon mass spectrum \cite{2+1glueball} can be found at arbitrarily small, but anisotropic gauge coupling.There is, unfortunately, a crossover from 
$(1+1)$-dimensional to $(2+1)$-dimensional behavior. A similar crossover
is an obstacle to using the form factors of the two-dimensional Ising spin field to 
calculate critical exponents of the three-dimensional Ising model. Konik and Adamov were able to overcome this dimensional crossover for the Ising case with a density-matrix real-space renormalization group \cite{KA}. The triviality of the {\em S} matrix as $N\rightarrow \infty$ may help defeat the crossover for SU($N$) gauge theories. The reason is that the energy eigenstates  of the ${\rm SU}(\infty)_{L}\times {\rm SU}(\infty)_{R}$
sigma model are simply Fock states of Bosons, in the appropriate basis. Our hope is that this will make a real-space-renormalization-group approach feasible for the non-Abelian gauge theory.

We assume no previous knowledge of exact form factors. The 
reader unfamiliar with 
integrable-bootstrap methods could simply take the $1/N$-expanded form of the 
{\em S} matrix (in Equation (\ref{expanded-s-matrix}) below) on faith. Otherwise, we recommend 
starting with the summary by Zamolodchikov and Zamolodchikov 
\cite{ZamX2}. The task of working through Reference \cite{ZamX2} may be simplified by consulting 
Reference \cite{SW}
(especially for infinite-product formulas for the {\em S} matrix) and
the appendix of the first of References \cite{FFgauge} (in which some results are derived from scratch). We also recommend Reference \cite{STW}, in which the sine law 
is explained. From there, the papers on U($N$)- and 
SU($N$)-invariant theories
of Berg et. al. \cite{BKKW} and Kurak and Swieca \cite{KS} should be accessible. With this preparation, the reader should be ready to follow the 
derivation of the {\em S} matrix of the principal chiral model 
\cite{AALS}, \cite{Wiegmann}. 

In the next section we discuss integrability and the $1/N$-expansion of the principal chiral model. We find the matrix element of the field operator between the vacuum and three-particle (more precisely one-antiparticle, two-antiparticle) state in Section 3. We write the leading terms of two-point Wightman function in Section 4. The form factors may be thought of as an expansion of the field operator in terms of a a free field, which we briefly discuss in Section 5. We present some conclusions and open questions in Section 6.

\section{The $1/N$-expansion of the {\em S} matrix and the field algebra}
\setcounter{equation}{0}
\renewcommand{\theequation}{2.\arabic{equation}}

The basic Wightman correlation function is
\beq
{\mathcal W}(x) = \frac{1}{N}\langle 0\vert {\rm Tr} \Phi(0) \Phi(x)^{\dagger} \vert 0\rangle, 
\label{wightman}
\eeq 
where the scaling field $\Phi$ is defined by (\ref{phi}) and the normalization condition
\beq
\langle 0\vert \Phi(0)_{b_{0} a_{0}}\vert P,\theta, a_{1}, b_{1}\rangle
=N^{-1/2}\delta_{a_{0} a_{1}}\delta_{b_{0} b_{1}}, \label{norm}
\eeq
where the ket on the right is a one particle ($r=1$) state, with rapidity $\theta$ (that is, with momentum components $p_{0}=m\cosh\theta$, $p_{1}=m\sinh\theta$) and we implicitly sum over 
left and right colors
$a_{1}$ and $b_{1}$, respectively.

The expression (\ref{norm}) is the most elementary form factor. It is similar to the definition of
the scaling field in the Ising model \cite{Ising}. We will 
determine the normalization of the other form factors using (\ref{norm}) 
and the LSZ reduction formula. The leading contribution to the Wightman function comes from the one-particle-intermediate-state approximation (or free-field approximation)
\beq
{\mathcal W}(x)\approx  \frac{1}{N}
\int \frac{d\theta}{4\pi}\, e^{{\rm i}m(x^{0}\cosh \theta- x^{1}\sinh \theta) }\,
\langle 0\vert \Phi(0)_{b_{0} a_{0}}\vert P,\theta, a_{1}, b_{1}\rangle_{\rm in}\,_{\rm in}
\langle P,\theta, a_{1}, b_{1} \vert \Phi(0)_{b_{0} a_{0}}^{*} \vert 0 \rangle, \label{1p}
\eeq
where $m$ denotes $m_{1}$ and the sum over all repeated color indices is implicit. For 
$x^{0}=0$, $x^{1}=\pm\vert x\vert$, this is
\beq
{\mathcal W}(x)
\approx \frac{1}{4\pi}K_{0}(m\vert x\vert). \nonumber
\eeq

Note that this expression is of order 
$(1/N)^{0}$. We are assuming that there is no contribution from the one-antiparticle state (with
$r=N-1$), {\em i.e.}
\beq
\langle 0\vert \Phi(0)_{b_{0} a_{0}}\vert A,\theta, b_{1}, a_{1}\rangle_{\rm in}
=0. \nonumber
\eeq

The 
{\em S} matrix can be determined, assuming unitarity, factorization (the Yang-Baxter relation) and maximal analyticity. The basic $r=1$ excitations have two color indices from $1$ to $N$. One can view these excitations as a bound pair of two quarks of different color sectors (or alternatively as a quark
in one color sector and an antiquark in the other). Such quarks can be regarded 
as the elementary physical 
excitations of the chiral Gross-Neveu model \cite{BKKW}, \cite{KS}, \cite{AL}. 

Next we show the {\em S} matrix of two elementary particles of the sigma model, with incoming
rapidities $\theta_{1}$ and 
$\theta_{2}$ (we use the definition $(p_{j})_{0}=m\cosh\theta_{j}$,
$(p_{j})_{1}=m\sinh \theta_{j}$, relating the momentum vector $p_{j}$ and rapidity $\theta_{j}$),
outgoing rapidities
$\theta_{1}^{\prime}$ and $\theta_{2}^{\prime}$
and rapidity difference $\theta=\theta_{12}=\theta_{1}-\theta_{2}$. This is
\begin{eqnarray}
{\mathcal S}_{PP}=S_{PP}(\vert \theta\vert)\;
4\pi \delta(\theta_{1}^{\prime}-\theta_{1})\;
4\pi \delta(\theta_{2}^{\prime}-\theta_{2}),   \nonumber
\end{eqnarray}
where $S_{PP}(\vert \theta \vert)$ is a function which acts on the quantum numbers of
the particles (in some papers, $\delta(p_{j}-p_{j}^{\prime})$ is written, incorrectly, in place of
$4\pi \delta(\theta_{j}-\theta_{j}^{\prime})$). The quantity $S_{PP}(\vert \theta \vert)$ is nearly always referred to as the {\em S} matrix
in the literature. It is
explicitly given by
\begin{eqnarray}
S_{PP}(\theta)
=
\frac{\sin (\theta/2-\pi{\rm i}/N)}{\sin(\theta/2+\pi{\rm i}/N)}\;S_{\rm CGN}(\theta)_{L}\otimes 
S_{\rm CGN}(\theta)_{R} ,\label{s-matrix}
\end{eqnarray}
where $S_{\rm CGN}(\theta)_{L,R}$, for either the subscript L (left) or R (right), is the {\em S} matrix of two elementary excitations of the chiral Gross-Neveu 
model:
\begin{eqnarray}
S_{\rm CGN}(\theta)\!\!=\!\!\frac{\Gamma({\rm i}\theta/2\pi+1)\Gamma(-{\rm i}\theta/2\pi-1/N) }{\Gamma({\rm i}\theta/2\pi+1-1/N) \Gamma(-{\rm i}\theta/2\pi)}
\left( 1-\frac{2\pi{\rm i}}{N\theta}P\right), \label{CGN}
\end{eqnarray}
where $P$ switches the 
colors of the elementary Gross-Neveu particles. {\em S} matrix elements for which one or both 
particles have $r>1$ can be found by fusion.

We shall define the generalized {\em S} matrix to be (\ref{s-matrix}) with $\vert\theta\vert$ replaced by
$\theta=\theta_{12}=\theta_{1}-\theta_{2}$. This is consistent with the definition given in Reference 
\cite{BabKarow} (where
it is called the auxillary {\em S} matrix).

The first few terms of the $1/N$-expansion of (\ref{s-matrix}) are \cite{Wiegmann}
\beq
S_{PP} (\theta)
=\left[ 1+O(1/N^{2})\right]
\left[1-\frac{2\pi {\rm i}}{N\theta}(P\otimes 1+1\otimes P)-\frac{4\pi^{2}}{N^{2}\theta^{2}}P\otimes P
\right]. \label{expanded-s-matrix}
\eeq
We can find the scattering matrix of one particle and one antiparticle $S_{AP}(\theta)$ from
(\ref{expanded-s-matrix}), using crossing.

There is are exceptional values of $\theta$ where the particle-particle {\em S} matrix does not become unity as $N\rightarrow \infty$. One of these is at $\theta=0$. For
vanishing relative rapidity, equation (\ref{s-matrix}) yields 
$S_{PP}(0)=-P\otimes P$, independently of $N$; thus the
expansion (\ref{expanded-s-matrix}) is not valid at $\theta=0$. A similar breakdown of the $1/N$-expansion at $\theta=0$ occurs for models with O($N$) symmetry 
\cite{ZamX2}, \cite{Z2}. This point corresponds to the threshold $s=4m^{2}$, where $s$ is the Mandelstam variable, related to the relative rapidity by $s=2m^{2}+2m^{2}\cosh \theta$. At this threshold, both particles have vanishing momenta in the center-of-mass frame, and exchange their left and right colors with probability one. In relativistic scattering theory
the {\em S} matrix has a cut from the $s$-channel threshold 
$s=4m^{2}$ to $s=\infty$, and another cut from the $t$-channel threshold $s=4m^{2}-t=0$ to $s=-\infty$. Another exceptional value where the {\em S} matrix is not unity as $N\rightarrow \infty$ is 
$\theta=2\pi{\rm i}/N$, where the $r=2$ bound state occurs. In the complex
$\theta$-plane, the first cut is the image of the line ${\rm Im}\;\theta=0$, and the other cut is
the image of the line ${\rm Im}\;\theta=\pi$ \cite{ZamX2}. Between these two lines, in the interior 
of the so-called physical strip, excluding bound-state poles, the expansion (\ref{expanded-s-matrix}) is valid, which is sufficient for the remaining
discussion in this paper.

The basic properties of particle states is encoded in the Zamolodchikov 
algebra. Let us introduce particle-creation operators ${\mathfrak A}^{\dagger}_{P}(\theta)_{ab}$ and
antiparticle-creation operators ${\mathfrak A}^{\dagger}_{A}(\theta)_{ba}$. This algebra
is essentially a non-Abelian particle-statistics relation:
\beq
{\mathfrak A}^{\dagger}_{P}(\theta_{1})_{a_{1}b_{1}}\,
{\mathfrak A}^{\dagger}_{P}(\theta_{2})_{a_{2}b_{2}}
&=&S_{PP}(\theta_{12})^{c_{2}d_{2};c_{1}d_{1}}_{a_{1}b_{1};a_{2}b_{2}}\;
{\mathfrak A}^{\dagger}_{P}(\theta_{2})_{c_{2}d_{2}}\,
{\mathfrak A}^{\dagger}_{P}(\theta_{1})_{c_{1}d_{1}} \nonumber \\
{\mathfrak A}^{\dagger}_{A}(\theta_{1})_{b_{1}a_{1}}\,
{\mathfrak A}^{\dagger}_{A}(\theta_{2})_{b_{2}a_{2}}
&=&S_{AA}(\theta_{12})^{d_{2}c_{2};d_{1}c_{1}}_{b_{1}a_{1};b_{2}a_{2}}\;
{\mathfrak A}^{\dagger}_{A}(\theta_{2})_{d_{2}c_{2}}\,
{\mathfrak A}^{\dagger}_{A}(\theta_{1})_{d_{1}c_{1}} \nonumber \\
{\mathfrak A}^{\dagger}_{P}(\theta_{1})_{a_{1}b_{1}}\,
{\mathfrak A}^{\dagger}_{A}(\theta_{2})_{b_{2}a_{2}}
&=&S_{AP}(\theta_{12})^{d_{2}c_{2};c_{1}d_{1}}_{a_{1}b_{1};b_{2}a_{2}}\;
{\mathfrak A}^{\dagger}_{A}(\theta_{2})_{d_{2}c_{2}}\,
{\mathfrak A}^{\dagger}_{P}(\theta_{1})_{c_{1}d_{2}}\,.
\label{creation}
\eeq
The Yang-Baxter relation is necessary as a consistency condition for 
(\ref{creation}). That is one way to understand why
the absence of
particle production implies integrability.

An in-state is defined as a product of creation operators in the order of increasing rapidity, from right to left, acting on the vacuum, {\em e.g.}
\beq
\vert P,\theta_{1},a_{1},b_{1};A,\theta_{2},b_{2},a_{2},\dots \rangle_{\rm in}
={\mathfrak A}^{\dagger}_{P}(\theta_{1})_{a_{1}b_{1}}
{\mathfrak A}^{\dagger}_{A}(\theta_{2})_{b_{2}a_{2}}\cdots \vert 0\rangle,\;\;{\rm where}\;
\theta_{1}>\theta_{2}>\cdots
\eeq
Similarly, an out-state is a product of creation operators in the order of decreasing rapidity, from right to left, acting on the vacuum.

The expression (\ref{expanded-s-matrix}) becomes unity as $N\rightarrow \infty$, as we would expect. The algebra
(\ref{creation})
thereby trivializes. Consider the field 
\beq
M(x)=\int \frac{d\theta}{4\pi}\, \left[{\mathfrak A}_{P}(\theta)
e^{{\rm i}m x^{0}\cosh \theta-{\rm i}m 
x^{1}\sinh \theta} 
+{\mathfrak A}^{\dagger}_{A}(\theta)e^{-{\rm i}m x^{0}\cosh \theta+{\rm i}m x^{1}\sinh \theta} 
\right], \label{master}
\eeq
where ${\mathfrak A}_{A}$ is the destruction operator of an antiparticle. It is simply the adjoint of the operator  ${\mathfrak A}^{\dagger}_{A}$. In the limit $N\rightarrow \infty$, $[{\mathfrak A}_{A,P}(\theta),
{\mathfrak A}^{\dagger}_{A,P}(\theta)]\rightarrow 4\pi \delta(\theta-\theta^{\prime})$, with all other commutators approaching zero (the commutators are more complicated for finite$N$). The $N\times N$-matrix-valued field operator $M(x)$ is a massive free field. The form factors give the coefficients of an expansion of 
the renormalized field $\Phi(x)$ in terms of this field.

The form factors are matrix elements between the vacuum and multi-particle in-states of the field
operator $\Phi$. The action of the global-symmetry transformation on $\Phi$ and the creation operators is
\beq
\Phi(x)\rightarrow V_{L}\Phi(x) V_{R}, \;\;
{\mathfrak A}^{\dagger}_{P}(\theta)\rightarrow V_{R}^{\dagger}
{\mathfrak A}^{\dagger}_{P}(\theta)V_{L}^{\dagger},\;\;
{\mathfrak A}^{\dagger}_{A}(\theta)\rightarrow V_{L}
{\mathfrak A}^{\dagger}_{P}(\theta)V_{R}. \label{transformation-laws}
\eeq
Thus we expect that, for large $N$, the condition
\beq
\langle 0\vert \Phi(0) \vert \Psi \rangle\neq 0,
\nonumber
\eeq
on an in-state $\vert \Psi\rangle$, which is an eigenstate of particle number, holds only if
$\vert \Psi\rangle$ contains $m$ particles and $m-1$ antiparticles, for some $m=1,2,\dots$. In the next section, we will find these matrix elements for $m=2$ (the $m=1$ case has already been discussed above).

\section{Maximally-analytic form factors}
\setcounter{equation}{0}
\renewcommand{\theequation}{3.\arabic{equation}}

In this section we will study matrix elements of the form $\langle 0\vert \Phi(0) \vert \Psi \rangle$, where
$\vert \Psi\rangle$ is an in-state with two elementary particles and one 
antiparticle, {\em i.e.} $m=2$. This matrix element is defined for general choices of
rapidity. Here are the form factors corresponding to different orderings of rapidities:
\beq
\langle 0 \vert\, \Phi(0)_{b_{0}a_{0}}\, \vert A,\theta_{1}, b_{1},a_{1}; P,\theta_{2}, a_{2},b_{2};
P,\theta_{3}, a_{3},b_{3}\rangle_{\rm in} 
&=&
\langle 0\vert \, \Phi(0)_{b_{0}a_{0}}\,\, {\mathfrak A}^{\dagger}_{A}(\theta_{1})_{b_{1}a_{1}}
{\mathfrak A}^{\dagger}_{P}(\theta_{2})_{a_{2}b_{2}}
{\mathfrak A}^{\dagger}_{P}(\theta_{3})_{a_{3}b_{3}}\,\vert 0\rangle
\nonumber \\
=\frac{1}{N^{3/2}}F_{1}(\theta_{1}, \theta_{2}, \theta_{3}) 
\delta_{a_{0}a_{2}}\delta_{b_{0}b_{3}} \delta_{b_{1}b_{2}}\delta_{a_{1}a_{3}}
&+&\frac{1}{N^{3/2}}F_{2}(\theta_{1}, \theta_{2}, \theta_{3}) 
\delta_{a_{0}a_{3}}\delta_{b_{0}b_{2}}\delta_{a_{1}a_{2}}\delta_{b_{1}b_{3}}\nonumber \\
+\frac{1}{N^{3/2}}F_{3}(\theta_{1}, \theta_{2}, \theta_{3}) 
\delta_{a_{0}a_{2}}\delta_{b_{0}b_{2}}\delta_{a_{1}a_{3}}\delta_{b_{1}b_{3}}
&+&\frac{1}{N^{3/2}}F_{4}(\theta_{1}, \theta_{2}, \theta_{3}) 
\delta_{a_{0}a_{3}}\delta_{b_{0}b_{3}}\delta_{b_{1}b_{2}}\delta_{a_{1}a_{2}}, 
\label{F-form-factors}
\eeq
for $\theta_{1}>\theta_{2}>\theta_{3}$,
\beq
\langle 0 \vert\, \Phi(0)_{b_{0}a_{0}}\, \vert P,\theta_{1}, a_{1},b_{1}; A,\theta_{2}, b_{2},a_{2};
P,\theta_{3}, a_{3},b_{3}\rangle_{\rm in}  
&=&
\langle 0\vert \, \Phi(0)_{b_{0}a_{0}}\,\, {\mathfrak A}^{\dagger}_{P}(\theta_{2})_{a_{2}b_{2}}
{\mathfrak A}^{\dagger}_{A}(\theta_{1})_{b_{1}a_{1}}
{\mathfrak A}^{\dagger}_{P}(\theta_{3})_{a_{3}b_{3}}\,\vert 0\rangle
\nonumber \\
=\frac{1}{N^{3/2}}{\tilde F}_{1}(\theta_{1}, \theta_{2}, \theta_{3}) 
\delta_{a_{0}a_{2}}\delta_{b_{0}b_{3}} \delta_{b_{1}b_{2}}\delta_{a_{1}a_{3}}
&+&\frac{1}{N^{3/2}}{\tilde F}_{2}(\theta_{1}, \theta_{2}, \theta_{3}) 
\delta_{a_{0}a_{3}}\delta_{b_{0}b_{2}}\delta_{a_{1}a_{2}}\delta_{b_{1}b_{3}}\nonumber \\
+\frac{1}{N^{3/2}}{\tilde F}_{3}(\theta_{1}, \theta_{2}, \theta_{3}) 
\delta_{a_{0}a_{2}}\delta_{b_{0}b_{2}}\delta_{a_{1}a_{3}}\delta_{b_{1}b_{3}}
&+&\frac{1}{N^{3/2}}{\tilde F}_{4}(\theta_{1}, \theta_{2}, \theta_{3}) 
\delta_{a_{0}a_{3}}\delta_{b_{0}b_{3}} \delta_{b_{1}b_{2}}\delta_{a_{2}a_{1}},
\label{tilde-F}
\eeq
for $\theta_{2}>\theta_{1}>\theta_{3}$, and
\beq
\langle 0 \vert\, \Phi(0)_{b_{0}a_{0}}\, \vert P,\theta_{1}, a_{1},b_{1}; P,\theta_{2}, a_{2},b_{2};
A,\theta_{3}, b_{3},a_{3}\rangle_{\rm in}  
&=&
\langle 0\vert \, \Phi(0)_{b_{0}a_{0}}\,\, {\mathfrak A}^{\dagger}_{P}(\theta_{2})_{a_{2}b_{2}}
{\mathfrak A}^{\dagger}_{P}(\theta_{3})_{a_{3}b_{3}}
{\mathfrak A}^{\dagger}_{A}(\theta_{1})_{b_{1}a_{1}}\,\vert 0\rangle
\nonumber \\
=\frac{1}{N^{3/2}}{\tilde{\tilde F}}_{1}(\theta_{1}, \theta_{2}, \theta_{3}) 
\delta_{a_{0}a_{2}}\delta_{b_{0}b_{3}} \delta_{b_{1}b_{2}}\delta_{a_{1}a_{3}}
&+&\frac{1}{N^{3/2}}{\tilde{\tilde F}}_{2}(\theta_{1}, \theta_{2}, \theta_{3}) 
\delta_{a_{0}a_{3}}\delta_{b_{0}b_{2}}\delta_{a_{1}a_{2}}\delta_{b_{1}b_{3}}\nonumber \\
+\frac{1}{N^{3/2}}{\tilde{\tilde F}}_{3}(\theta_{1}, \theta_{2}, \theta_{3}) 
\delta_{a_{0}a_{2}}\delta_{b_{0}b_{2}}\delta_{a_{1}a_{3}}\delta_{b_{1}b_{3}}
&+&\frac{1}{N^{3/2}}{\tilde{\tilde F}}_{4}(\theta_{1}, \theta_{2}, \theta_{3}) 
\delta_{a_{0}a_{3}}\delta_{b_{0}b_{3}} \delta_{b_{1}b_{2}}\delta_{a_{2}a_{1}},
\label{tilde-tilde-F}
\eeq
for $\theta_{3}>\theta_{1}>\theta_{2}$. We note that (\ref{F-form-factors}) is equivalent to
\beq
\langle 0 \vert\, \Phi(0)_{b_{0}a_{0}}\, \vert A,\theta_{1}, b_{1},a_{1}; P,\theta_{3}, a_{3},b_{3};
P,\theta_{2}, a_{2},b_{2}\rangle_{\rm in}  
&=&
\langle 0\vert \, \Phi(0)_{b_{0}a_{0}}\,\, {\mathfrak A}^{\dagger}_{A}(\theta_{1})_{b_{1}a_{1}}
{\mathfrak A}^{\dagger}_{P}(\theta_{3})_{a_{3}b_{3}}
{\mathfrak A}^{\dagger}_{P}(\theta_{2})_{a_{2}b_{2}}\,\vert 0\rangle
\nonumber \\
=\frac{1}{N^{3/2}}F_{2}(\theta_{1}, \theta_{3}, \theta_{2}) 
\delta_{a_{0}a_{2}}\delta_{b_{0}b_{3}} \delta_{b_{1}b_{2}}\delta_{a_{1}a_{3}}
&+&\frac{1}{N^{3/2}}F_{1}(\theta_{1}, \theta_{3}, \theta_{2}) 
\delta_{a_{0}a_{3}}\delta_{b_{0}b_{2}}\delta_{a_{1}a_{2}}\delta_{b_{1}b_{3}}\nonumber \\
+\frac{1}{N^{3/2}}F_{4}(\theta_{1}, \theta_{3}, \theta_{2}) 
\delta_{a_{0}a_{2}}\delta_{b_{0}b_{2}}\delta_{a_{1}a_{3}}\delta_{b_{1}b_{3}}
&+&\frac{1}{N^{3/2}}F_{3}(\theta_{1}, \theta_{3}, \theta_{2}) 
\delta_{a_{0}a_{3}}\delta_{b_{0}b_{3}}\delta_{b_{1}b_{2}}\delta_{a_{1}a_{2}}, 
\label{hat-F-form-factors}
\eeq
for $\theta_{1}>\theta_{3}>\theta_{2}$.

We generalize the form factor \cite{BabKarow}, so that (\ref{F-form-factors}), (\ref{tilde-F}), (\ref{tilde-tilde-F}) and (\ref{hat-F-form-factors}) are valid without the inequalities
on the arguments $\theta_{1,2,3}$.

In each of the expressions (\ref{F-form-factors}), (\ref{tilde-F}), (\ref{tilde-tilde-F}) and (\ref{hat-F-form-factors}), we have written the quantity on the right in a similar way. Each of the products of Kronecker deltas is a possible covariant tensor of the global color symmetry. No other combinations are allowed for $N>3$, by equation (\ref{transformation-laws}).


Notice that Lorentz invariance implies that the scalar functions $F$, $G$ and $H$ are unchanged under an overall boost $\theta_{j}\rightarrow \theta_{j}+\Delta \theta$, $j=1,2,3$. This means that the form factors depend only on differences of the rapidities.

If we examine the contribution of these form factors to the Wightman function $C(x)$, defined in (\ref{wightman}), we see that $F$, $\tilde F$ and $\tilde {\tilde F}$ must be multiplied by 
$N^{-3/2}$, as we have in (\ref{F-form-factors}), (\ref{tilde-F}), (\ref{tilde-tilde-F}) and (\ref{hat-F-form-factors}). We will eventually show in this section that $F_{3,4}$, ${\tilde F}_{3,4}$ and 
${\tilde {\tilde F}}_{3,4}$ are down by a further power of $N$. This means we could have written 
(\ref{F-form-factors}), (\ref{tilde-F}), (\ref{tilde-tilde-F}) and (\ref{hat-F-form-factors}) with the coefficient $1/N^{5/2}$ in front of the last two entries, instead of
$1/N^{3/2}$. These are the coefficients of
tensors where the both quantum numbers of the antiparticle coincide with both of those for the one of the particles. For the time being, however, we will
treat $F_{3,4}$, ${\tilde F}_{3,4}$ and ${\tilde {\tilde F}}_{3,4}$ just like the other functions.


First we apply the scattering form-factor axiom, also called Watson's theorem. This axiom can be most simply understood as the application of  the Zamolodchikov algebra to the vacuum expectation values in the first lines of equations
(\ref{F-form-factors}), (\ref{tilde-F}) and (\ref{tilde-tilde-F}) above. It is essentially the assumption that we can continue the functions $F$, $G$ and $H$ outside the domain 
$\theta_{1}<\theta_{2}<\theta_{3}$, in such a way that the Zamolodchikov algebra is satisfied. For example, if we apply Watson's theorem on the incoming antiparticle with rapidity $\theta_{1}$ and the incoming
particle with rapidity $\theta_{2}$, on the left-hand side of (\ref{F-form-factors}) we find
\beq
\langle0\vert \, \Phi(0)_{b_{0}a_{0}}\,\, 
{\mathfrak A}^{\dagger}_{P}(\theta_{1})_{a_{1}b_{1}}
{\mathfrak A}^{\dagger}_{A}(\theta_{2})_{b_{2}a_{2}}
{\mathfrak A}^{\dagger}_{P}(\theta_{3})_{a_{3}b_{3}}\,\vert 0\rangle
=S_{AP}(\theta_{12})^{d_{2}c_{2};c_{1}d_{1}}_{a_{1}b_{1};b_{2}a_{2}}
\langle 0\vert \, \Phi(0)_{b_{0}a_{0}}\,\, 
{\mathfrak A}^{\dagger}_{A}(\theta_{2})_{d_{2}c_{2}}
{\mathfrak A}^{\dagger}_{P}(\theta_{1})_{c_{1}d_{1}}
{\mathfrak A}^{\dagger}_{P}(\theta_{3})_{a_{3}b_{3}}\, \vert 0\rangle
. \label{watson-1}
\eeq
The $1/N$-expansion of the {\em S} matrix element in (\ref{watson-1}) is
\beq
&S_{AP}\!\!\!&\!\!\!(\theta_{12})^{d_{2}c_{2};c_{1}d_{1}}_{a_{1}b_{1};b_{2}a_{2}}=\left[ 1+O(1/N^{2})\right] \nonumber \\
&\times& \left[
\delta^{d_{2}}_{b_{2}}\delta^{c_{2}}_{a_{2}}\delta^{c_{1}}_{a_{1}}\delta^{d_{1}}_{b_{1}}
-\frac{2\pi{\rm i}}{N{\hat \theta_{12}}}
\!\left(\! \delta_{a_{1}a_{2}}\delta^{c_{1}c_{2}}\delta^{d_{2}}_{b_{2}}\delta^{d_{1}}_{b_{1}}
+\delta^{c_{2}}_{a_{2}}\delta^{c_{1}}_{a_{1}}\delta_{b_{1}b_{2}}\delta^{d_{1}d_{2}}
\! \right)\! -\frac{4\pi^{2}}{N^{2}{\hat \theta}_{12}^{2}}
\delta_{a_{1}a_{2}}\delta^{c_{1}c_{2}}\delta_{b_{1}b_{2}}\delta^{d_{1}d_{2}} 
  \right], \label{crossed-S}
\eeq 
where ${\hat \theta_{12}}=\pi {\rm i}-\theta_{12}$ is the rapidity difference after crossing from the 
$s$-channel to the $t$-channel. Inserting the explicit expressions on the right-hand sides of 
(\ref{F-form-factors}) and (\ref{tilde-F}) into (\ref{watson-1})
and after some work, we find
\beq
{\tilde F}(\theta_{1},\theta_{2},\theta_{3})=\left(
\begin{array}{cccc}
1-\frac{2\pi{\rm i}}{{\hat \theta}_{12}} \;\;&\;\; 0\;\; & \;\;-\frac{2\pi{\rm i}}{N{\hat \theta}_{12}}\;\;&\;\; 0\\
0&1-\frac{2\pi{\rm i}}{{\hat \theta}_{12}} & -\frac{2\pi{\rm i}}{N{\hat \theta}_{12}}&0\\
0&0&1&0\\
-\frac{2\pi{\rm i}}{N{\hat \theta}_{12}}(1-\frac{2\pi{\rm i}}{{\hat \theta}_{12}})&
-\frac{2\pi{\rm i}}{N{\hat \theta}_{12}}(1-\frac{2\pi{\rm i}}{{\hat \theta}_{12}})
&0&(1-\frac{2\pi{\rm i}}{{\hat \theta}_{12}})^{2}
\end{array}
\right)
F(\theta_{1},\theta_{2},\theta_{3}) +O\left( \frac{1}{N^{2}}\right), \label{reduced-watson-1}
\eeq
where we have denoted the four-component vectors in the obvious way, {\em e.g.}
\beq
F(\theta_{1},\theta_{2},\theta_{3})
=\left( \begin{array}{c}
F_{1}(\theta_{1}, \theta_{2}, \theta_{3}) \\
F_{2}(\theta_{1}, \theta_{2}, \theta_{3}) \\
F_{3}(\theta_{1}, \theta_{2}, \theta_{3}) \\
F_{4}(\theta_{1}, \theta_{2}, \theta_{3}) \end{array}
\right) .\nonumber
\eeq


In finding (\ref{reduced-watson-1}) some factors of $N$ appeared as a result of contracting indices. These factors of $N$ canceled some factors
of $1/N$ in the second and third terms of the {\em S} matrix element in (\ref{crossed-S}).

There are two more useful relations following from the scattering axiom. These are 
\beq
\langle 0\vert \Phi(0)_{b_{0}a_{0}}
{\mathfrak A}^{\dagger}_{P}(\theta_{2})_{a_{2}b_{2}}
{\mathfrak A}^{\dagger}_{P}(\theta_{3})_{a_{3}b_{3}}
{\mathfrak A}^{\dagger}_{A}(\theta_{1})_{b_{1}a_{1}} \vert 0\rangle 
=S_{AP}(\theta_{13})^{d_{1}c_{1};c_{3}d_{3}}_{a_{3}b_{3};b_{1}a_{1}}
\langle 0\vert \Phi(0)_{b_{0}a_{0}}
{\mathfrak A}^{\dagger}_{P}(\theta_{2})_{a_{2}b_{2}}
{\mathfrak A}^{\dagger}_{A}(\theta_{1})_{d_{1}c_{1}}
{\mathfrak A}^{\dagger}_{P}(\theta_{3})_{c_{3}d_{3}}
 \vert 0 \rangle, \nonumber
\eeq
which may be re-expressed as
\beq
{\tilde{\tilde F}}(\theta_{1},\theta_{2},\theta_{3})=
\left(
\begin{array}{cccc}
1-\frac{2\pi{\rm i}}{{\hat \theta}_{13}} \;\;&\;\; 0\;\; & \;\;0\;\;&\;\; -\frac{2\pi{\rm i}}{N{\hat \theta}_{13}}\\
0&1-\frac{2\pi{\rm i}}{{\hat \theta}_{13}} & 0&-\frac{2\pi{\rm i}}{N{\hat \theta}_{13}}\\
-\frac{2\pi{\rm i}}{N{\hat \theta}_{13}}(1-\frac{2\pi{\rm i}}{{\hat \theta}_{13}})&
-\frac{2\pi{\rm i}}{N{\hat \theta}_{13}}(1-\frac{2\pi{\rm i}}{{\hat \theta}_{13}})
&(1-\frac{2\pi{\rm i}}{{\hat \theta}_{13}})^{2}&0\\
0&0&0&1
\end{array}
\right)
{\tilde F}(\theta_{1},\theta_{2} ,\theta_{3})+O\left( \frac{1}{N^{2}}\right), \label{reduced-watson-2}
\eeq
and finally
\beq
\langle 0\vert \Phi(0)_{b_{0}a_{0}}
{\mathfrak A}^{\dagger}_{A}(\theta_{1})_{b_{1}a_{1}}
{\mathfrak A}^{\dagger}_{P}(\theta_{2})_{a_{2}b_{2}}
{\mathfrak A}^{\dagger}_{P}(\theta_{3})_{a_{3}b_{3}} \vert 0 \rangle
=S_{PP}(\theta_{23})^{c_{2}d_{2};c_{3}d_{3}}_{a_{2}b_{2};a_{3}b_{3}}
\langle 0\vert \Phi(0)_{b_{0}a_{0}}
{\mathfrak A}^{\dagger}_{A}(\theta_{1})_{b_{1}a_{1}}
{\mathfrak A}^{\dagger}_{P}(\theta_{3})_{c_{3}d_{3}}
{\mathfrak A}^{\dagger}_{P}(\theta_{2})_{c_{2}d_{2}}
 \vert 0 \rangle, \nonumber
\eeq
which reduces to
\beq
F(\theta_{1}, \theta_{2}, \theta_{3}) =\left(
\begin{array}{cccc}
0\;\;& \;\;1\;\;&\;\; -\frac{2\pi{\rm i}}{N\theta_{23}}\;\;&\;\;-\frac{2\pi{\rm i}}{N\theta_{23}}\\
1&0& -\frac{2\pi{\rm i}}{N\theta_{23}}&-\frac{2\pi{\rm i}}{N\theta_{23}}\\
-\frac{2\pi{\rm i}}{N\theta_{23}}& -\frac{2\pi{\rm i}}{N\theta_{23}}&0 &1\\
-\frac{2\pi{\rm i}}{N\theta_{23}}& -\frac{2\pi{\rm i}}{N\theta_{23}}&1 &0
\end{array}
\right)
F(\theta_{1},\theta_{3},\theta_{2}) +O\left( \frac{1}{N^{2}}\right). \label{reduced-watson-3}
\eeq

Now in (\ref{reduced-watson-2}), some factors of $1/N$ in {\em S} matrix elements were canceled after summing over indices, as we noted above for (\ref{reduced-watson-1}). This did not happen in obtaining (\ref{reduced-watson-3}). The reason is that the particle-particle {\em S} matrix 
(\ref{expanded-s-matrix}) does not
contract colors of incoming particles; colors can only be exchanged.

Another of Smirnov's axioms is the periodicity condition. This axiom is an application of 
crossing. Explicitly:
\beq
\langle 0\vert \Phi(0)_{b_{0}a_{0}}\;{\mathfrak A}^{\dagger}_{I_{1}}(\theta_{1})_{C_{1}} 
{\mathfrak A}^{\dagger}_{I_{2}}(\theta_{2})_{C_{2}} \cdots
{\mathfrak A}^{\dagger}_{I_{M}}(\theta_{M})_{C_{M}} \vert 0\rangle 
=\langle 0\vert \Phi(0)_{b_{0}a_{0}}\;
{\mathfrak A}^{\dagger}_{I_{M}}(\theta_{M}-2\pi{\rm i})_{C_{M}} \;
{\mathfrak A}^{\dagger}_{I_{1}}(\theta_{1})_{C_{1}} \cdots
{\mathfrak A}^{\dagger}_{I_{M-1}}(\theta_{M-1})_{C_{M-1}} \vert 0\rangle ,
\label{axiom2}
\eeq
where $I_{k}$, $k=1,\dots,M$ is $P$ or $A$ (particle or antiparticle) and $C_{k}$ denotes a pair
of indices (which may be written $a_{k}b_{k}$, for $C_{k}=P$ and
 $b_{k}a_{k}$, for $C_{k}=A$).  A brief explanation of (\ref{axiom2}) follows. For more details, see  Reference \cite{BabKarow}. Consider what happens when a creation operator
in front of the ket is replaced by an annihilation operator behind the bra 
by crossing. Consider the vacuum expectation value of creation operators and 
$\Phi(0)_{b_{0}a_{0}}$
\beq
\langle \!\! \!\!\!\!&0\!\!\!\!&\!\! \vert {\mathfrak A}_{I_{1}}(\theta_{1})_{C_{1}}\; \Phi(0)_{b_{0}a_{0}}\;
{\mathfrak A}^{\dagger}_{I_{M}}(\theta_{M})_{C_{M}}
{\mathfrak A}^{\dagger}_{I_{M-1}}(\theta_{M-1})_{C_{M-1}}
\cdots
{\mathfrak A}^{\dagger}_{I_{2}}(\theta_{2})_{C_{2}} \vert 0\rangle_{\rm connected}  \nonumber \\ 
&\!\!=\!\!& \langle0\vert {\mathfrak A}_{I_{1}}(\theta_{1})_{C_{1}}\; \Phi(0)_{b_{0}a_{0}}\;
{\mathfrak A}^{\dagger}_{I_{M}}(\theta_{M})_{C_{M}}
{\mathfrak A}^{\dagger}_{I_{M-1}}(\theta_{M-1})_{C_{M-1}}
\cdots
{\mathfrak A}^{\dagger}_{I_{2}}(\theta_{2})_{C_{2}}\vert 0\rangle \nonumber \\
&\!\!-\!\!& \langle0\vert {\mathfrak A}_{I_{1}}(\theta_{1})_{C_{1}}\; \Phi(0)_{b_{0}a_{0}}\vert 0\rangle
\langle 0 \vert
{\mathfrak A}^{\dagger}_{I_{M}}(\theta_{M})_{C_{M}}
{\mathfrak A}^{\dagger}_{I_{M-1}}(\theta_{M-1})_{C_{M-1}}
\cdots
{\mathfrak A}^{\dagger}_{I_{2}}(\theta_{2})_{C_{2}}\vert 0\rangle \;.\nonumber
\eeq
The subscript ``connected" is included because the vacuum intermediate channel is subtracted 
\cite{BabKarow}. This expression means $M-1$ incoming particles 
are absorbed by a ``probe", corresponding to the operator $\Phi(0)_{b_{0}a_{0}}$. This probe then
emits a single particle. Consider the pair of particles, with labels $1$ (the outgoing particle)
and $M$. Under crossing, these both become incoming particles, but with $\theta_{1}$
replaced by $\theta_{1}-\pi {\rm i}$. The reason is that $\theta_{1}\rightarrow \theta_{1}-\pi{\rm i}$ preserves
the relativistic invariants $s_{j\;j+1}=(p_{j}+ p_{j+1})^{2}$, and
$t_{j\;j+1}=(p_{j}- p_{j+1})^{2}$, where $j=2,\dots,M-1$, while
interchanging the two invariants $s_{1M}=(p_{1}+p_{M})^{2}$ and 
$t_{1M}=(p_{1}-p_{M})^{2}$. Thus
\beq
\langle \!\!\!  \!\!\!&0\!\!\!\!&\!\! \vert {\mathfrak A}_{I_{1}}(\theta_{1})_{C_{1}}\; \Phi(0)_{b_{0}a_{0}}\;
{\mathfrak A}_{I_{M}}(\theta_{M})_{C_{M}}^{\dagger}{\mathfrak A}_{I_{M-1}}(\theta_{M-1})_{C_{M-1}}^{\dagger}
\cdots
{\mathfrak A}_{I_{2}}(\theta_{2})_{C_{2}}^{\dagger} \vert 0\rangle_{\rm connected}  \nonumber \\ 
&\!\!=\!\!&
\langle 0\vert \Phi(0)_{b_{0}a_{0}} {\mathfrak A}^{\dagger}_{I_{1}}(\theta_{1}-\pi{\rm i})_{C_{1}}
{\mathfrak A}^{\dagger}_{I_{2}}(\theta_{2})_{C_{2}}
\cdots {\mathfrak A}^{\dagger}_{I_{M}}(\theta_{M})_{C_{M}}\vert 0\rangle
\;.
\label{crossing1}
\eeq
Suppose that instead of interchanging the invariants $s_{1M}$ and $t_{1M}$, we interchange
the invariants $s_{12}=(p_{1}+p_{2})^{2}$ and $t_{12}=(p_{1}-p_{2})^{2}$. Then we find
\beq
\langle \!\!\!  \!\!\!&0\!\!\!\!&\!\! \vert {\mathfrak A}_{I_{1}}(\theta_{1})_{C_{1}}\; \Phi(0)_{b_{0}a_{0}}\;
{\mathfrak A}_{I_{M}}(\theta_{M})_{C_{M}}^{\dagger}{\mathfrak A}_{I_{M-1}}(\theta_{M-1})_{C_{M-1}}^{\dagger}
\cdots
{\mathfrak A}_{I_{2}}(\theta_{2})_{C_{2}}^{\dagger} \vert 0\rangle_{\rm connected}  \nonumber \\ 
&\!\!=\!\!&
\langle 0\vert \Phi(0)_{b_{0}a_{0}} {\mathfrak A}^{\dagger}_{I_{2}}(\theta_{2})_{C_{2}}
{\mathfrak A}^{\dagger}_{I_{3}}(\theta_{3})_{C_{3}}
\cdots {\mathfrak A}^{\dagger}_{I_{M}}(\theta_{M})_{C_{M}}
 {\mathfrak A}^{\dagger}_{I_{1}}(\theta_{1}+\pi{\rm i})_{C_{1}}
\vert 0\rangle
\;.
\label{crossing2}
\eeq
The periodicity axiom  (\ref{axiom2}) follows from 
(\ref{crossing1}) and (\ref{crossing2}).

Notice that integrability was not used to justify
(\ref{axiom2}). The periodicity axiom follows from very 
general considerations in $1+1$ dimensions \cite{Nied}.

The periodicity axiom implies the three relations
\beq
\langle 0 \vert \Phi(0)_{b_{0}a_{0}}
{\mathfrak A}^{\dagger}_{A}(\theta_{1}-2\pi{\rm i})_{b_{1}a_{1}}
{\mathfrak A}^{\dagger}_{P}(\theta_{2})_{a_{2}b_{2}}
{\mathfrak A}^{\dagger}_{P}(\theta_{3})_{a_{3}b_{3}} \vert 0\rangle
&=&\langle 0 \vert \Phi(0)_{b_{0}a_{0}}
{\mathfrak A}^{\dagger}_{P}(\theta_{2})_{a_{2}b_{2}}
{\mathfrak A}^{\dagger}_{P}(\theta_{3})_{a_{3}b_{3}} 
{\mathfrak A}^{\dagger}_{A}(\theta_{1})_{b_{1}a_{1}}\vert 0\rangle,
\nonumber \\
\langle 0 \vert \Phi(0)_{b_{0}a_{0}}
{\mathfrak A}^{\dagger}_{P}(\theta_{2}-2\pi{\rm i})_{a_{2}b_{2}}
{\mathfrak A}^{\dagger}_{A}(\theta_{1})_{b_{1}a_{1}}
{\mathfrak A}^{\dagger}_{P}(\theta_{3})_{a_{3}b_{3}} \vert 0\rangle
&=&\langle 0 \vert \Phi(0)_{b_{0}a_{0}}
{\mathfrak A}^{\dagger}_{A}(\theta_{1})_{b_{1}a_{1}}
{\mathfrak A}^{\dagger}_{P}(\theta_{3})_{a_{3}b_{3}} 
{\mathfrak A}^{\dagger}_{P}(\theta_{2})_{a_{2}b_{2}}\vert 0\rangle,
\nonumber \\
\langle 0 \vert \Phi(0)_{b_{0}a_{0}}
{\mathfrak A}^{\dagger}_{P}(\theta_{2}-2\pi{\rm i})_{a_{2}b_{2}}
{\mathfrak A}^{\dagger}_{P}(\theta_{3})_{a_{3}b_{3}}
{\mathfrak A}^{\dagger}_{A}(\theta_{1})_{b_{1}a_{1}} \vert 0\rangle
&=&\langle 0 \vert \Phi(0)_{b_{0}a_{0}}
{\mathfrak A}^{\dagger}_{P}(\theta_{3})_{a_{3}b_{3}}
{\mathfrak A}^{\dagger}_{A}(\theta_{1})_{b_{1}a_{1}} 
{\mathfrak A}^{\dagger}_{P}(\theta_{2})_{a_{2}b_{2}}\vert 0\rangle,
\nonumber 
\eeq
which may be written as
\beq
F(\theta_{1}-2\pi{\rm i}, \theta_{2}, \theta_{3})
={\tilde {\tilde F}}(\theta_{1}, \theta_{2}, \theta_{3}),
\label{periodicity1} \\
{\tilde F}(\theta_{1}, \theta_{2}-2\pi{\rm i}, \theta_{3})
=F(\theta_{1}, \theta_{3}, \theta_{2}),
\label{periodicity2} \\
{\tilde {\tilde F}}(\theta_{1}, \theta_{2}-2\pi{\rm i}, \theta_{3})
= {\tilde F}(\theta_{1}, \theta_{3}, \theta_{2}),
\label{periodicity3}
\eeq
respectively.

Our work is simplified by expanding the form factors in powers of $1/N$:
\beq
F(\theta_{1},\theta_{2},\theta_{3})
=
F^{0}(\theta_{1},\theta_{2},\theta_{3})+\frac{1}{N}F^{1}(\theta_{1},\theta_{2},\theta_{3})
+\cdots \;,
\label{FFexp}
\eeq
and similarly for ${\tilde F}(\theta_{1},\theta_{2},\theta_{3})$ and
${\tilde{\tilde F}}(\theta_{1},\theta_{2},\theta_{3})$. We truncate this expansion to leading order, keeping only 
$F^{0}(\theta_{1},\theta_{2},\theta_{3})$, ${\tilde F}^{0}(\theta_{1},\theta_{2},\theta_{3})$ and
${\tilde{\tilde F}}^{0}(\theta_{1},\theta_{2},\theta_{3})$.

Combining (\ref{reduced-watson-1}) and (\ref{reduced-watson-2})
with (\ref{periodicity1}), we find
\beq
F^{0}_{1}(\theta_{1}-2\pi{\rm i},\theta_{2}, \theta_{3})&=&
\frac{\theta_{12}+\pi{\rm i}}{\theta_{12}-\pi{\rm i}}
\frac{\theta_{13}+\pi{\rm i}}{\theta_{13}-\pi{\rm i}}
\;F^{0}_{1}(\theta_{1},\theta_{2}, \theta_{3})\; ,\nonumber \\
F^{0}_{2}(\theta_{1}-2\pi{\rm i},\theta_{2}, \theta_{3})&=&
\frac{\theta_{12}+\pi{\rm i}}{\theta_{12}-\pi{\rm i}}
\frac{\theta_{13}+\pi{\rm i}}{\theta_{13}-\pi{\rm i}}
\;F^{0}_{2}(\theta_{1},\theta_{2}, \theta_{3})\;,\nonumber \\
F^{0}_{3}(\theta_{1}-2\pi{\rm i},\theta_{2}, \theta_{3})&=&
\left( \frac{\theta_{13}+\pi{\rm i}}{\theta_{13}-\pi{\rm i}}\right)^{2}
\;F^{0}_{3}(\theta_{1},\theta_{2}, \theta_{3})\;,\nonumber \\
F^{0}_{4}(\theta_{1}-2\pi{\rm i},\theta_{2}, \theta_{3})&=&
\left( \frac{\theta_{12}+\pi{\rm i}}{\theta_{12}-\pi{\rm i}} \right)^{2}
\;F^{0}_{4}(\theta_{1},\theta_{2}, \theta_{3})\; .\label{reduced-periodicity} 
\eeq
Thus the components of the form factor are periodic, except for 
phases. Furthermore, (\ref{reduced-watson-3}) implies that
\beq
F^{0}_{1}(\theta_{1},\theta_{2},\theta_{3})
=F^{0}_{2}(\theta_{1},\theta_{3},\theta_{2})\;,\;\;
F^{0}_{3}(\theta_{1},\theta_{2},\theta_{3})
=F^{0}_{4}(\theta_{1},\theta_{3},\theta_{2}). \label{switching}
\eeq

The general solution of (\ref{reduced-periodicity}) and (\ref{switching}) is
\beq
F^{0}_{1}(\theta_{1},\theta_{2}, \theta_{3}) &=&
(\theta_{12}+\pi{\rm i})^{-1}(\theta_{13}+\pi{\rm i})^{-1} g_{1}( \theta_{1}, \theta_{2},\theta_{3})\;,\nonumber\\
F^{0}_{2}(\theta_{1},\theta_{2}, \theta_{3}) &=&
(\theta_{12}+\pi{\rm i})^{-1}(\theta_{13}+\pi{\rm i})^{-1} g_{1}( \theta_{1}, \theta_{3},\theta_{2})\;,\nonumber\\
F^{0}_{3}(\theta_{1},\theta_{2}, \theta_{3}) &=&
({\theta_{13}+\pi{\rm i}})^{-2}
\;g_{3}( \theta_{1}, \theta_{2},\theta_{3})\;,\nonumber\\
F^{0}_{4}(\theta_{1},\theta_{2}, \theta_{3}) &=&
 (\theta_{12}+\pi{\rm i})^{-2}\;g_{3}( \theta_{1}, \theta_{3},\theta_{2})\;, \label{gensoln}
\eeq
where the functions $g_{1}$ and $g_{3}$ are periodic in $\theta_{1}$:
\beq
g_{1}( \theta_{1}-2\pi{\rm i}, \theta_{2},\theta_{3})=g_{1}( \theta_{1}, \theta_{2},\theta_{3})\;,\;\;
g_{3}( \theta_{1}-2\pi{\rm i}, \theta_{2},\theta_{3})=g_{3}( \theta_{1}, \theta_{2},\theta_{3})\;.
\nonumber
\eeq

We now turn to the remaining periodicity conditions (\ref{periodicity2}) and (\ref{periodicity3}). Combining (\ref{reduced-watson-2}) with (\ref{periodicity2}), we find
\beq
\frac{\theta_{12}+3\pi{\rm i}}{\theta_{12}+\pi{\rm i}}\;
F^{0}_{1,2}(\theta_{1},\theta_{2}-2\pi{\rm i},\theta_{3})
=F^{0}_{2,1}(\theta_{1},\theta_{2},\theta_{3}) &,& \nonumber \\
F^{0}_{3}(\theta_{1},\theta_{2}-2\pi{\rm i},\theta_{3})=
F^{0}_{4}(\theta_{1},\theta_{2},\theta_{3})&,& \nonumber \\
\left(\frac{\theta_{12}+3\pi{\rm i}}{\theta_{12}+\pi{\rm i}}\right)^{2}\;
F^{0}_{4}(\theta_{1},\theta_{2}-2\pi{\rm i},\theta_{3})=
F^{0}_{3}(\theta_{1},\theta_{2},\theta_{3})
&,& \label{lastperiod1}
\eeq
and combining (\ref{reduced-watson-1}) and
(\ref{reduced-watson-2}) with (\ref{periodicity3}) yields 
\beq
\frac{\theta_{12}+3\pi{\rm i}}{\theta_{12}+\pi{\rm i}}\frac{\theta_{13}+\pi{\rm i}}{\theta_{13}-\pi{\rm i}}
\; F^{0}_{1,2}(\theta_{1},\theta_{2}-2\pi{\rm i},\theta_{3})
&=&\frac{\theta_{13}+\pi{\rm i}}{\theta_{13}-\pi{\rm i}}\; 
F^{0}_{2,1}(\theta_{1},\theta_{3},\theta_{2})\;,
\nonumber \\
\left(\frac{\theta_{13}+3\pi{\rm i}}{\theta_{13}+\pi{\rm i}}\right)^{2}\; 
F^{0}_{3}(\theta_{1},\theta_{2}-2\pi{\rm i},\theta_{3})&=& 
\left( \frac{\theta_{12}+\pi{\rm i}}{\theta_{12}-\pi{\rm i}} \right)^{2}\; 
F^{0}_{4}(\theta_{1},\theta_{2},\theta_{3})\;, \nonumber \\
\left(\frac{\theta_{12}+3\pi{\rm i}}{\theta_{12}+\pi{\rm i}}\right)^{2}\;
F^{0}_{4}(\theta_{1},\theta_{2}-2\pi{\rm i},\theta_{3})&=&
F^{0}_{3}(\theta_{1},\theta_{2},\theta_{3})
\;. \label{lastperiod2}
\eeq
The first of (\ref{lastperiod1}) and the first of (\ref{lastperiod2}) are the same equation. The last of 
(\ref{lastperiod1}) and the last of (\ref{lastperiod2}) are the same equation. The second of
(\ref{lastperiod1}) and the second of (\ref{lastperiod2}) are inconsistent unless
\beq
F^{0}_{3}(\theta_{1},\theta_{2},\theta_{3})=F^{0}_{4}(\theta_{1},\theta_{2},\theta_{3})=0, 
\label{vanishing}
\eeq
which we claimed at the beginning of this section. Thus the double poles 
in (\ref{gensoln}) are absent. 
The conditions (\ref{lastperiod1}) and
(\ref{lastperiod2}) imply
\beq
g_{1}(\theta_{1},\theta_{2}-2\pi{\rm i},\theta_{3})=g_{1}(\theta_{1},\theta_{3},\theta_{2}).
\nonumber 
\eeq

The minimal choice of the form factor, with no unnecessary poles or zeros, satisfying both
Watson's theorem and the periodicity axiom, is obtained
by setting the function $g_{1}( \theta_{1}, \theta_{3},\theta_{2})$ equal to a constant:
\beq
F^{0}_{1}(\theta_{1},\theta_{2}, \theta_{3}) =
\frac{g_{1}}{(\theta_{12}+\pi{\rm i})(\theta_{13}+\pi{\rm i}) }\;, 
\;\;
F^{0}_{2}(\theta_{1},\theta_{2}, \theta_{3}) =
\frac{g_{1}}{(\theta_{12}+\pi{\rm i})(\theta_{13}+\pi{\rm i})}\;.\nonumber
\eeq
We fix the constant with the annihilation-pole axiom.

The annihilation-pole axiom concerns the residues of form factors at singularities. This axiom follows from the LSZ reduction formula. The derivation can be found in
Reference \cite{BabKarow}, but some clarification may be helpful to the reader. We take the 
field $\Phi$ in the left-hand side of (\ref{F-form-factors})
on the mass shell, and compare with the {\em S} matrix. We first cross the antiparticle: 
$\theta_{1}
\rightarrow \theta_{1}-\pi{\rm i}$. So now we are considering two particles, of rapidities $\theta_{2}$
and $\theta_{3}$, in the initial state. These scatter and there is a particle (not antiparticle) of
rapidity $\theta_{1}$ in the final state. There must also be a second particle in the final state, which corresponds to taking $\Phi$ on shell; we denote its rapidity by $\theta_{0}$. The reduction formula
is
\beq
\;_{\rm out}\langle \!\!\!\!\!&\!\!\! P\!\!\!&\!\!\!\!\!,\theta_{1}, a_{1},b_{1};P,\theta_{0}, a_{0},b_{0}
\vert P,\theta_{2}, a_{2},b_{2};
P,\theta_{3}, a_{3},b_{3}
\rangle_{\rm in}
=\;_{\rm out}\langle P,\theta_{1}, a_{1},b_{1}\vert P,\theta_{2}, a_{2},b_{2}\rangle_{\rm in}
\;_{\rm out}\langle P,\theta_{0}, a_{0},b_{0}\vert P,\theta_{3}, a_{3},b_{3}\rangle_{\rm in}
\nonumber \\
 &+&\;_{\rm out}\langle P,\theta_{1}, a_{1},b_{1}\vert P,\theta_{3}, a_{3},b_{3}\rangle_{\rm in}
\;_{\rm out}\langle P,\theta_{0}, a_{0},b_{0}\vert P,\theta_{2}, a_{2},b_{2}\rangle_{\rm in}
\nonumber \\
&+&{\rm i}{\sqrt{N}}\int d^{2}x\;e^{{\rm i}mx^{0}\cosh\theta_{0}-{\rm i}mx^{1}\sinh \theta_{0}} 
\;_{\rm out}\langle P,\theta_{1}, a_{1},
b_{1}\vert 
\;(\partial_{0}^{2}-\partial_{1}^{2}+m^{2})\Phi(x)_{b_{0}a_{0}} \;\vert P, \theta_{2}, a_{2}, b_{2};
P, \theta_{3}, a_{3}, b_{3}
\rangle_{\rm in}, \label{lsz}
\eeq
where the factor $\sqrt N$ comes 
from the normalization of $\Phi$
(\ref{1p}). The second term on the right-hand 
side of (\ref{lsz}) vanishes if $\theta_{1}<\theta_{0}$. The right-hand side is the particle-particle 
{\em S} matrix element, which can be directly compared with (\ref{expanded-s-matrix}).

To evaluate the right-hand side of (\ref{lsz}), we use the the formulas
for the Klein-Gordon operator \cite{Ising}
\beq
(p_{1}-p_{2}-p_{3})^{2}-m^{2}=-8m^{2} \sinh\frac{\theta_{12}}{2}
\sinh\frac{\theta_{13}}{2}\cosh\frac{\theta_{23}}{2} , \label{KGop}
\eeq
and for the covariant delta function
\beq
\delta^{2}(p_{1}+{p_{0}-p_{2}}-p_{3})
&=&\delta[(p_{1})_{+} +(p_{0})_{+} -(p_{2})_{+} -(p_{3})_{+} ]\;
\delta[(p_{1})_{-} +(p_{0})_{-} -(p_{2})_{-} -(p_{3})_{-} ]\nonumber \\
&=&\frac{2}{m^{2}}\delta[(p_{1})_{+} +(p_{0})_{+} -(p_{2})_{+} -(p_{3})_{+} ]\;
\delta[(p_{1})_{+}^{-1} +(p_{0})_{+}^{-1} -(p_{2})_{+}^{-1} -(p_{3})_{+}^{-1} ]\nonumber \\
&=&\frac{2}{m^{2}} \left\vert \frac{1}{(p_{3})_{+}^{2}}-\frac{1}{(p_{2})_{+}^{2}}
\right\vert^{-1} \delta[(p_{1})_{+}-(p_{2})_{+}]\;\delta[(p_{3})_{+}-(p_{0})_{+}] \nonumber \\
&+&\frac{2}{m^{2}} \left\vert \frac{1}{(p_{3})_{+}^{2}}-\frac{1}{(p_{2})_{+}^{2}}
\right\vert^{-1} \delta[(p_{1})_{+}-(p_{3})_{+}]\;\delta[(p_{2})_{+}-(p_{0})_{+}] \nonumber \\
&=&
\frac{\delta(\theta_{12})\delta(\theta_{30})}{m^{2}\vert \sinh\theta_{13}\vert}
+\frac{\delta(\theta_{13})\delta(\theta_{20})}{m^{2}\vert \sinh\theta_{12}\vert}, \label{cov-delta}
\eeq
where the components of each of the momenta along the light cone are 
$p_{\pm}=2^{-1/2}(p_{0}\pm p_{1})=2^{-1/2}e^{\pm \theta}$. We hope the indices cause no confusion; we have written $(p_{i})_{\mu}$ for the $\mu^{\rm th}$ component of the momentum of the $i^{\rm th}$ particle.

Inserting (\ref{KGop}) and (\ref{cov-delta}) into (\ref{lsz}), finally crossing the out-particle with rapidity 
$\theta_{1}$ back to an in-antiparticle with $\theta_{1}\rightarrow \theta_{1}+\pi{\rm i}$,  gives the annihilation-pole axiom for the problem in this section. Explicitly:
\beq
{\rm  Res}\vert_{\theta_{12}=-\pi{\rm i}} 
\langle 0 \vert\, \Phi(0)_{b_{0}a_{0}}\, \vert A,\theta_{1}, b_{1},a_{1}; P,\theta_{2}, a_{2},b_{2};
P,\theta_{3}, a_{3},b_{3}\rangle 
=2{\rm i} \langle 0 \vert\, \Phi(0)_{b_{0}a_{0}}\, \vert 
P,\theta_{3}, a_{3},b_{3}\rangle 
\left[\delta_{a_{1}a_{2}}\delta_{b_{1}b_{2}} -S^{a_{1}b_{1}}_{a_{2}b_{2}}(\theta_{23})\right], 
\nonumber \\
{\rm  Res}\vert_{\theta_{13}=-\pi{\rm i}} 
\langle 0 \vert\, \Phi(0)_{b_{0}a_{0}}\, \vert A,\theta_{1}, b_{1},a_{1}; P,\theta_{2}, a_{2},b_{2};
P,\theta_{3}, a_{3},b_{3}\rangle 
=2{\rm i} \langle 0 \vert\, \Phi(0)_{b_{0}a_{0}}\, \vert 
P,\theta_{2}, a_{2},b_{2}\rangle 
\left[\delta_{a_{1}a_{3}}\delta_{b_{1}b_{3}} -S^{a_{1}b_{1}}_{a_{3}b_{3}}(\theta_{23})\right].
\label{a-p}
\eeq
The leading terms of each side of (\ref{a-p}) are both of order $N^{-3/2}$. Our
final Lorentz-invariant expression for the large-$N$ limit of the one-antiparticle, two-particle form factor is
\beq
F^{0}_{1}(\theta_{1},\theta_{2}, \theta_{3}) &=&F^{0}_{2}(\theta_{1},\theta_{2}, \theta_{3}) \;=\;
-\frac{4\pi}{(\theta_{12}+\pi{\rm i})(\theta_{13}+\pi{\rm i})}\;,\nonumber \\
F^{0}_{3}(\theta_{1},\theta_{2}, \theta_{3})&=&F^{0}_{4}(\theta_{1},\theta_{2}, \theta_{3})\;=\;0.
\label{3part}
\eeq
The other functions ${\tilde F}^{0}_{j}(\theta_{1},\theta_{2}, \theta_{3})$ and 
${\tilde {\tilde F}}^{0}_{j}(\theta_{1},\theta_{2}, \theta_{3})$ are the same as 
$F^{0}_{j}(\theta_{1},\theta_{2}, \theta_{3})$, up to irrelevant phases (these phases disappear upon
evaluation of Wightman functions).

\section{The Wightman function in the 't~Hooft limit}
\setcounter{equation}{0}
\renewcommand{\theequation}{4.\arabic{equation}}

We can use the result of the previous section to find an improved expression for the $N=\infty$ two-point Wightman function (\ref{wightman}):
\beq
{\mathcal W}(x)&=&  \frac{1}{N}
\int \frac{d\theta}{4\pi}\, e^{{\rm i}m(x^{0}\cosh \theta- x^{1}\sinh \theta) }\,
\langle 0\vert \;\Phi(0)_{b_{0} a_{0}}\;\vert P,\theta, a_{1}, b_{1}\rangle_{\rm in}\,_{\rm in}
\langle P,\theta, a_{1}, b_{1} \vert \;\Phi(0)_{b_{0} a_{0}}^{*} \;\vert 0 \rangle \nonumber \\
&+& \frac{1}{N}
\int \frac{d\theta_{1}}{4\pi} \int \frac{d\theta_{2}}{4\pi}
\int \frac{d\theta_{3}}{4\pi}\;\frac{1}{2!}\;e^{{\rm i}m \sum_{j=1}^{3}(x^{0}\cosh \theta_{j}- x^{1}\sinh \theta_{j}) }\,
\langle 0\vert  \;  \Phi(0)_{b_{0}a_{0}} \; \vert
A, \theta_{1},a_{1},b_{1}; P,\theta_{2}, a_{2}, b_{2};
P,\theta_{3} , a_{3},b_{3}
\rangle_{\rm in}
\nonumber \\
&\times&\,_{\rm in}\langle A, \theta_{1},a_{1},b_{1}; P,\theta_{2}, a_{2}, b_{2};
P,\theta_{3} , a_{3},b_{3}
\vert \;\Phi(0)_{b_{0} a_{0}}^{*} \;\vert 0 \rangle\; +\;\cdots \;,
 \label{2p}
\eeq
where, as in (\ref{1p}), we sum over repeated color indices.

All of the one-antiparticle, two-particle form factors are given by (\ref{3part}) up to an 
irrelevant phase. When summing over color indices, we find that contributions quadratic in
either 
$F_{1}^{0}$ or $F_{2}^{0}$ are of order one. The mixed contributions, linear in both $F_{1}^{0}$ and 
$F_{2}^{0}$ are down by a power of $1/N$. We therefore drop the latter contributions.  Thus the expansion (\ref{2p}) is 
\beq
{\mathcal W}(x)
=\frac{1}{4\pi}\int d\theta\,e^{{\rm i}m(x^{0}\cosh \theta- x^{1}\sinh \theta) }
+\frac{1}{4\pi}\int d^{3}\theta\;e^{{\rm i}m \sum_{j=1}^{3}(x^{0}\cosh \theta_{j}- x^{1}\sinh \theta_{j}) }\,
(\theta_{12}^{2}+\pi^{2})^{-1}(\theta_{13}^{2}+\pi^{2})^{-1}+\cdots \;. \label{Wightman-exp}
\eeq
The first term on the right-hand side is the 
free-field approximation, discussed in Section 2. The result 
(\ref{Wightman-exp}) should be extremely good at large distances, as contributions from more intermediate particles fall off more quickly. Unfortunately, we cannot recover the short-distance behavior predicted by perturbation theory. It is necessary to sum over all intermediate states to 
obtain the Wightman function for small $x$. In other words, all the form factors of 
$\Phi$ are needed to
compare with the perturbative result.

\section{The correspondence with a free field}
\setcounter{equation}{0}
\renewcommand{\theequation}{5.\arabic{equation}}

The renormalized field can be written in terms of the Zamolodchikov particle-creation operators, and
their adjoints (together these form the Faddeev-Zamolodchikov algebra, which we do not discuss here). At large $N$, these are the standard operators used to build a free complex $(\infty\times \infty)$-matrix field $M(x)$ in (\ref{master}).

Examining the definitions of the functions $F$, $\tilde F$ and $\tilde{\tilde F}$ gives an expansion for $\Phi(x)$:
\beq
\Phi(x)_{b_{0}a_{0}}&=&\frac{1}{N^{1/2}}M(x)_{b_{0}a_{0}}
\nonumber \\
&-&\frac{1}{N^{3/2}}\int \frac{d^{3}\theta}{(4\pi)^{3}}
[{\mathfrak A}_{A}(\theta_{1})_{a_{1}b_{1}}
e^{{\rm i}m x^{0}\cosh \theta_{1}-{\rm i}m 
x^{1}\sinh \theta_{1}} +{\mathfrak A}_{P}^{\dagger}(\theta_{1})_{a_{1}b_{1}}
e^{-{\rm i}m x^{0}\cosh \theta_{1}+{\rm i}m 
x^{1}\sinh \theta_{1}} ] \nonumber \\
&\times &  \frac{1}{2!}[{\mathfrak A}_{P}(\theta_{2})_{b_{2}a_{2}}
e^{{\rm i}m x^{0}\cosh \theta_{2}-{\rm i}m 
x^{1}\sinh \theta_{2}} +{\mathfrak A}_{A}^{\dagger}(\theta_{2})_{b_{2}a_{2}}
e^{-{\rm i}m x^{0}\cosh \theta_{2}+{\rm i}m 
x^{1}\sinh \theta_{2}} ]  \nonumber \\
&\times &[{\mathfrak A}_{P}(\theta_{3})_{b_{3}a_{3}}
e^{{\rm i}m x^{0}\cosh \theta_{3}-{\rm i}m 
x^{1}\sinh \theta_{3}} +{\mathfrak A}_{A}^{\dagger}(\theta_{3})_{b_{3}a_{3}}
e^{-{\rm i}m x^{0}\cosh \theta_{3}+{\rm i}m 
x^{1}\sinh \theta_{3}} ]  \nonumber \\
&\times & \frac{4\pi}{(\theta_{12}+\pi{\rm i})(\theta_{13}+\pi{\rm i})}
\left(\frac{\theta_{12}+\pi{\rm i}}{\theta_{12}-\pi{\rm i}}
\right)^{\Theta(\theta_{12})}
\left(\frac{\theta_{13}+\pi{\rm i}}{\theta_{13}-\pi{\rm i}}
\right)^{\Theta(\theta_{13})}
(\delta_{a_{0}a_{2}} \delta_{b_{0}b_{3}} \delta_{a_{1}a_{3}} \delta_{b_{1}b_{2}}
+\delta_{a_{0}a_{3}} \delta_{b_{0}b_{2}} \delta_{a_{1}a_{2}} \delta_{b_{1}b_{3}}
)\nonumber \\
&+& \cdots
\;,
\label{master-expansion}
\eeq
where $\Theta$ is the step function, $\Theta(\theta)=0$, for $\theta<0$, and $\Theta(\theta)=1$, for 
$\theta>0$, and the operators $\mathfrak A$ and ${\mathfrak A}^{\dagger}$ are expressed in terms of
the free field as
\beq
{\mathfrak A}^{\dagger}_{A}(\theta)_{ba}&=&
(2m{\rm i}\cosh \theta)^{-1} \int dx^{1}\; e^{{\rm i}m x^{0}\cosh \theta_{1}-{\rm i}m 
x^{1}\sinh \theta_{1}} \;{\overleftrightarrow \partial_{0}} \;M(x)_{ba} \nonumber \\
{\mathfrak A}^{\dagger}_{P}(\theta)_{ab}&=&
(2m{\rm i}\cosh \theta)^{-1} \int dx^{1}\; e^{{\rm i}m x^{0}\cosh \theta_{1}-{\rm i}m 
x^{1}\sinh \theta_{1}} \;{\overleftrightarrow \partial_{0}} \;[M(x)^{\dagger}]_{ab}\;,
\eeq
and their adjoints. The matrix elements of this expression (\ref{master-expansion}) between the vacuum bra and an in-state ket are unchanged 
if we suppress the creation operators. The creation operators are needed, however,
for matrix elements of (\ref{master-expansion}) to satisfy crossing. 

\section{Conclusions}
\setcounter{equation}{0}
\renewcommand{\theequation}{6.\arabic{equation}}

To summarize, we found exact form factors for the (1+1)-dimensional
principal chiral model at large $N$. We expanded the two-point Wightman function in terms of these form factors. Finally, we identified an underlying free matrix field operator $M(x)$, and discussed how the renormalized field can be obtained from $M(x)$.

The $1/N$-expansion of the principal chiral model is quite different from the expansion of vector models, such as the O($N$) sigma model. The renormalized field of a vector model is a free 
field, as 
$N\rightarrow \infty$. 

There is little difference between the free massive field $M(x)$ and the classical master field of
the large-$N$ limit. The response of this field to a source is the same, whether or not it is quantized.

The ingredients to find higher-order corrections in the $1/N$-expansion (\ref{FFexp}) are already in Section 3. This problem is under investigation.  

It would be interesting to understand form factors for in-states with more particles. The number of functions rapidly increases with more particles. Nonetheless, two-antiparticle, three-particle form 
factors seem possible to obtain. It may be that all the form factors can be found. This would yield the complete sum of planar diagrams and a direct comparison with perturbation theory could be made. 

We have not discussed operators other than the renormalized field in this paper. It seems possible to find the form factors of currents and the energy-momentum tensor by similar methods.

\begin{acknowledgments}

I thank Kim Splittorff for discussions concerning the behavior of the first nontrivial part of the
Wightman function. This work was supported in 
part by the National Science Foundation, under Grant No. 
PHY0855387, and by a grant from the PSC-CUNY. I would also like to thank the Galileo Galilei Institute for the opportunity to present some of these ideas at the workshop ``Large-N Gauge Theories".

\end{acknowledgments}

\end{document}